\documentclass[11pt,aps,floatfix,tightenlines]{revtex4}
\usepackage{epsfig}
\usepackage{graphicx}
\usepackage{graphics}
\usepackage{xspace}
\usepackage{amssymb}
\usepackage{amsmath}
\usepackage{latexsym}
\usepackage{natbib}
\usepackage{mathrsfs}
\usepackage{url}
\newcommand{\inieq}{\begin{eqnarray}}            
\newcommand{\fineq}{\end{eqnarray}}            
\newcommand{\diff}{{\rm\,d}}                    

\begin{document}
\title{Strange quark effects in electron and neutrino-nucleus 
quasi-elastic scattering } 

\author{Andrea Meucci} 
\author{Carlotta Giusti}
\author{Franco Davide Pacati }
\affiliation{Dipartimento di Fisica Nucleare e Teorica, 
Universit\`{a} di Pavia and \\
Istituto Nazionale di Fisica Nucleare, 
Sezione di Pavia, I-27100 Pavia, Italy}


\begin{abstract}
The role of the sea quarks to ground state nucleon 
properties with electroweak probes is discussed.
A relativistic Green's function approach to parity violating electron 
scattering and a distorted-wave 
impulse-approximation applied to charged- and neutral-current 
neutrino-nucleus quasi-elastic scattering are presented in view of the 
possible determination of the strangeness content of the nucleon.
\end{abstract}
 
\maketitle

\section{Introduction}
\label{sec.int}

The nucleon is a bound state of three valence quarks. However, a sea of virtual
$q\bar q$ pairs and gluons surrounds each valence quark and play an important
role at distance scales of the bound state, where the QCD coupling constant
is large and the effects of the color field cannot be calculated accurately. 
One simple way to probe the 
effects of the sea is to investigate whether strange quarks contribute to the 
static properties of the nucleon. The first evidence that the strange axial
form factor
$g_{\mathrm A}^{\mathrm s} = G _{\mathrm A}^{\mathrm s} (Q^2=0)$ is different
from zero and large was found at CERN by the EMC collaboration \cite{emc} 
in a measurement of deep inelastic scattering of polarized muons on 
polarized protons.  
In order to study the role of the strange quark to
the spin structure of the nucleon various reactions have been proposed. Here we
are interested in parity-violating (PV) electron scattering and neutrino-nucleus
scattering. These two kinds of reactions can give us complementary 
information about the contributions of the sea quarks
to the properties of the nucleon. While PV electron scattering  is essentially
sensitive to the electric and magnetic strangeness of the nucleon,
neutrino-induced reactions are primarily sensitive to the axial-vector form
factor of the nucleon. 

A number of PV electron scattering measurements have been carried out in recent
years. They are sensitive to the strangeness contribution by measuring the
helicity-dependent PV asymmetry
\inieq A_{\mathrm{PV}} = \frac{\diff\sigma_+ - \diff\sigma_-}
{\diff\sigma_+ + \diff\sigma_-}\ , \label{eq.a}
\fineq
where $\diff\sigma_{+(-)}$ is the cross section for incident
right(left)-handed electrons from unpolarized targets (usually protons). 
$A_{\mathrm{PV}}$ arises from the
interference between electromagnetic and weak processes and depends on the
electric (magnetic) form factors $G^{\gamma}_{\mathrm {E(M)}}$, their weak
counterparts $G^{Z}_{\mathrm {E(M)}}$, and the axial form factor as seen in 
electron scattering $G^{e}_{{\mathrm A}}$. It has been noted that the 
contribution of radiative correction 
must be calculated in order to allow a precise extraction of the
strange axial form factor $G^{s}_{{\mathrm A}}$ from a PV measurement of 
$G^{e}_{{\mathrm A}}$. This usually prevents from a final determination of 
$G^{s}_{{\mathrm A}}$ from this data.
The SAMPLE \cite{spa2} results at backward angle on proton and deuteron 
targets reported
results for $G_{\mathrm{M}}^{\mathrm s}$ and $G^{e}_{{\mathrm A}}$ at $Q^2
\simeq 0.1$ (GeV/$c)^2$.
The HAPPEX \cite{happex2}, A4 \cite{a42}, and G0 \cite{g02} results at 
forward angles 
provided a linear combination of
$G_{\mathrm{E}}^{\mathrm s}$ and $G_{\mathrm{M}}^{\mathrm s}$ over the 
range $0.1
\leq Q^2 \leq 1$ (GeV/$c)^2$, where the contribution from the axial term is
usually suppressed by kinematical conditions. 
Three independent measurements are needed to
extract $G_{\mathrm{E}}^{\mathrm s}$, $G_{\mathrm{M}}^{\mathrm s}$, and
$G^{e}_{{\mathrm A}}$ separately. The PV asymmetry from a spinless, isoscalar 
target, such as $^4$He, depends only on the electric form 
factors \cite{happex24} and
represents an interesting via to avoid the problem of the axial term.  

Neutrino reactions are a well-established alternative to PV electron scattering
and give us complementary information about the contributions of the sea quarks
to the properties of the nucleon. 
A measurement of $\nu (\bar\nu$)-proton elastic scattering at Brookhaven
National Laboratory (BNL) \cite{bnl} suggested a non-zero value for the strange 
axial-vector form factor of the 
nucleon. However, it has been shown in Ref. \cite{gar} that the BNL data 
cannot provide us
decisive conclusions about the strange form factors when also strange vector 
form factors are taken into account.
The FINeSSE \cite{fin} experiment at Fermi National Laboratory aims 
at performing a detailed investigation of the strangeness contribution to the 
proton spin via measurements of the ratio of neutral-current to the 
charged-current $\nu
(\bar\nu)N$ processes. When combined with the existing data on PV scattering, a
determination of the strange form factors in the range 
$0.25 \leq Q^2 \leq 0.75$ would have to be possible with an uncertainty at 
each point of 
$\simeq \pm 0.02$~\cite{pate}. Since a significant part of the event will be 
from
scattering on $^{12}$C, nuclear structure effects have to be clearly understood 
in order to
give a reliable interpretation of the data.

\section{PV asymmetry in inclusive electron scattering on nuclei}
\label{sec2}

The helicity asymmetry for the scattering of a polarized electron on a target
nucleus through an angle $\vartheta$
can be written from Eq. (\ref{eq.a}) as the ratio between the
PV and the parity-conserving (PC) cross section, i.e.,
\inieq
A =  A_0 \frac {v^{}_{\mathrm{L}}R_{\mathrm{L}}^{\mathrm{AV}} +
 v^{}_{\mathrm{T}}
R_{\mathrm{T}}^{\mathrm{AV}} + v_{\mathrm{T}}'R_{\mathrm{T}}^{\mathrm{VA}}} 
{v^{}_{\mathrm{L}}R_{\mathrm{L}} + v^{}_{\mathrm{T}}R_{\mathrm{T}}}\ ,
\label{eq.A}\fineq
where $A_0\simeq 1.799 \times 10^{-4}\ Q^2$ (GeV/$c)^{-2}$ is a scale factor.
The coefficients $v$ are derived from the lepton tensor components and are 
taken from 
Ref.~\cite{pv}. The response functions $R$ are given in terms of the 
components of
the hadron tensor and contain the interference between the electromagnetic and 
the weak neutral part of the current operator~\cite{pv}.
The single-particle electromagnetic part of the current is
\begin{eqnarray}
  j^{\mu} =  F_1 \gamma ^{\mu} + 
      i\frac {\kappa}{2M} F_2\sigma^{\mu\nu}q_{\nu}	  
	     \ .
	     \label{eq.el}
\end{eqnarray}
The single-particle current operator related to the weak neutral current is  
\begin{eqnarray}
  j^{\mu} =  F_1^{\mathrm V} \gamma ^{\mu} + 
      i\frac {\kappa}{2M} F_2^{ \mathrm V}\sigma^{\mu\nu}q_{\nu}	  
	     -G_{ \mathrm A}\gamma ^{\mu}\gamma ^{5}
	     \ .
	     \label{eq.nc}
\end{eqnarray}
The vector form factors $F_i^{\mathrm V}$ can be expressed in terms of the 
corresponding electromagnetic form factors for protons $(F_i^{\mathrm p})$ and 
neutrons $(F_i^{\mathrm n})$, plus a possible isoscalar strange-quark 
contribution $(F_i^{\mathrm s})$, i.e.,
\begin{eqnarray}
F_i^{ \mathrm V;\ p(n)} = \pm \left\{F_i^p - F_i^n\right\}/2 - 
2\sin^2{\theta_{\mathrm W}}F_i^{p(n)} - F_i^s/2\ , \label{eq.fnc}
\end{eqnarray}
where $+(-)$ stands for proton (neutron) knockout and $\theta_{\mathrm W}$ is 
the Weinberg angle $(\sin^2{\theta_{\mathrm W}} 
\simeq 0.2313)$. 
The strange vector form factors are taken as 
\begin{eqnarray}
F_1^{\mathrm s}(Q^2) =  \frac {(\rho^{\mathrm s} + 
\mu^{\mathrm s}) \tau}{(1+\tau) (1+Q^2/M_{\mathrm V}^2)^2} \  , \ \ 
F_2^{\mathrm s}(Q^2) =  \frac {\left(\mu^{\mathrm s}-\tau \rho^{\mathrm s}  
\right)}{(1+\tau) (1+Q^2/M_{\mathrm V}^2)^2} \ ,
\label{eq.sform}
\end{eqnarray}
where $\tau = Q^2/(4M_{\mathrm p}^2)$ and $M_{\mathrm V}$ =
0.843 GeV. The quantities $\mu^{\mathrm s}$ and $\rho^{\mathrm s}$ are related
to the strange magnetic moment and radius of the nucleus. 
The axial form factor is expressed as 
\begin{eqnarray}
G_{{\mathrm A}}(Q^2)  =  \frac {1}{2}  
\left(\pm g_{{\mathrm A}}^{\phantom{ }} - g_{{{\mathrm A}}}^s\right) G   \ , 
 \label{eq.gas} 
\end{eqnarray}
where $g_{\mathrm A} \simeq 1.26$, 
$g^{\mathrm s}_{\mathrm A}$ describes possible strange-quark contributions, and 
\begin{eqnarray}
G = (1+Q^2/M_{\mathrm A}^2)^{-2} \ . \label{eq.ga}
\end{eqnarray} 
The axial mass has been taken 
from Ref. \cite{bernard} as $M_{\mathrm A}$ = (1.026$\pm$0.021) GeV.

The inclusive PV electron scattering may be treated using the same 
relativistic 
approach which was already applied to the inclusive PC electron 
scattering \cite{ee} and to the inclusive 
quasi-elastic $\nu$($\bar\nu$)-nucleus scattering \cite{cc}. The components of
the nuclear response are written in terms 
of the single-particle optical-model Green's function, that is
based on a bi-orthogonal expansion in terms of the eigenfunctions of the 
non-Hermitian optical potential and of its Hermitian conjugate. As it is
discussed in Refs. \cite{pv,ee,cc}, the flux is preserved and 
final state
interactions (FSI) are treated 
in the inclusive reaction consistently with the exclusive one. 
\begin{figure}[h]
\begin{center}
\includegraphics[width=3.5in]{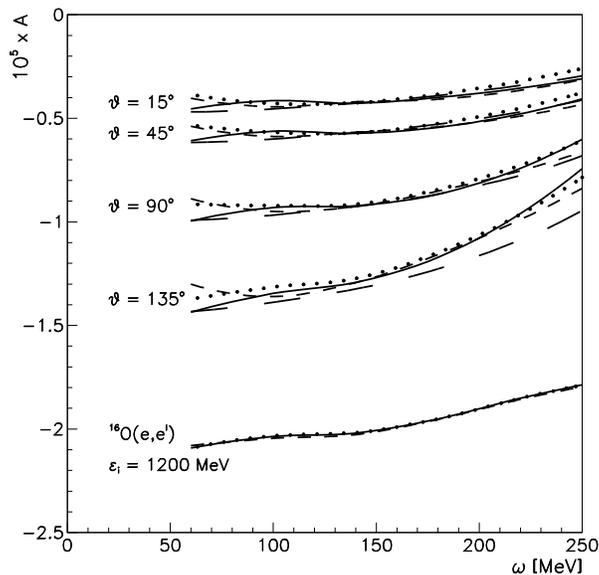} 
\vskip -0.3cm
\caption {PV asymmetry for $^{12}$C at $q$ = 400 MeV/c and $^{16}$O 
at $\varepsilon_i$ = 1200 MeV and $\vartheta = 32^{\mathrm o}$ with different 
bound states and optical potentials as explained in the text. The results are 
rescaled by the factor $10^5$. 
}
\label{fa3}
\end{center}
\end{figure}

In order to evaluate the uncertainties of the model, we compare in 
Fig.~\ref{fa3} the asymmetries for $^{12}$C at $q$ = 400 MeV and $^{16}$O at 
$\varepsilon_i$ = 1200 MeV and $\vartheta$ = 32$^{\mathrm o}$ 
calculated with different bound states and optical potentials.
The full lines give the results with the NL-SH bound states \cite{sharma} and 
the EDAD1 optical potentials \cite{chc}. 
The dotted lines give the same calculations, but with the NL3 bound states of 
Ref.~\cite{lala} and the EDAD1 optical potentials. The dashed lines are 
calculated with the NL-SH bound states and the 
energy-dependent and A-independent EDAI-12C or EDAI-16O optical 
potentials~\cite{chc}.
The differences are very small everywhere, but for $^{12}$C at low and high 
energy transfers, where they are in any case less than 10\%. 
The long dashed lines show the results for the RPWIA. 
Only small differences are found when they are compared with the Green's 
function results. This means that the ratio that gives the asymmetry 
cancels most of the effect of the FSI. 

\begin{figure}[h]
\begin{center}
\includegraphics[width=3.5in]{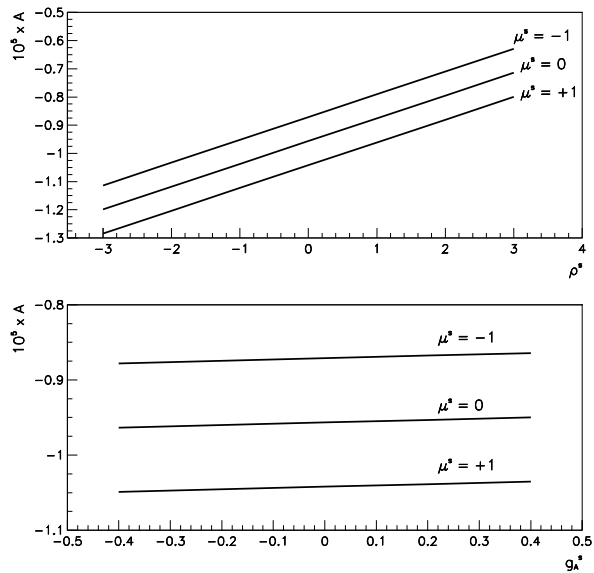} 
\vskip -0.3cm
\caption {PV asymmetry for $^{12}$C at $q$ = 500 MeV/c, 
$\omega$ = 120 MeV, and $\vartheta$ = 30$^{\mathrm o}$ as a function 
of $\rho^{\mathrm s}$ and $\mu^{\mathrm s}$ (upper panel) and as a function of 
$g_{\mathrm A}^{\mathrm s}$ and $\mu^{\mathrm s}$ (lower panel). 
}
\label{3d1}
\end{center}
\end{figure}

The sensitivity of PV electron scattering to the effect of 
strange-quark contribution to the vector and axial-vector form factors, is 
shown in Fig.~\ref{3d1} for $^{12}$C at 
$q$ = 500 MeV/c, $\omega$ = 120 MeV, and 
$\vartheta$ = 30$^{\mathrm o}$ as a function of 
the strangeness parameters, 
$\rho^{\mathrm s}$, $\mu^{\mathrm s}$, and $g^{\mathrm s}_{\mathrm A}$.  
The range of their values is chosen according to Refs.~\cite{happex2,beck1}. 
The asymmetry reduces in absolute value
up to $\simeq$40\% as $\rho^{\mathrm s}$ varies in the range  
$-$3 $\leq \rho^{\mathrm s}\leq $ $+$3, whereas it changes up 
to $\simeq$15\% for
$-$1 $\leq \mu^{\mathrm s}\leq $ $+$1. We note that, according to 
HAPPEX \cite{happex2} 
results,
$\rho^{\mathrm s}$ and $\mu^{\mathrm s}$ might have opposite sign, thus leading 
to a partial cancellation of the effects.
The sensitivity to $g^{\mathrm s}_{\mathrm A}$ is very weak, as can be seen  
in the
lower panel of Fig.~\ref{3d1}.

\section{The quasi-elastic neutrino-nucleus scattering}
\label{sec3}

The $\nu$($\bar\nu$)-nucleus cross section for the
semi-inclusive process may be 
written as a contraction between the lepton and the hadron 
tensor.
The lepton tensor is defined in a similar way as in electromagnetic knockout 
and it separates into a symmetrical and an anti-symmetrical component
which are written as in Refs. \cite{book,cc,nc}. 
The hadron tensor is given in its general 
form by suitable bilinear products of the 
transition matrix elements of the nuclear weak-current operator.
Assuming that the final states are
given by the product of a discrete (or continuum) state of the 
residual
nucleus and a scattering state of the emitted 
nucleon and using the impulse approximation, 
the transition amplitude reduces to 
the sum of terms similar to those appearing in the electron
scattering case \cite{book,meucci1}.
The single-particle current operator related to the weak current is  
\begin{eqnarray}
  j^{\mu} =  \Big[F_1^{\mathrm V} \gamma^{\mu} + 
             i\frac {\kappa}{2M} F_2^{\mathrm
	     V}\sigma^{\mu\nu}q_{\nu}	  
  -G_{\mathrm A}\gamma ^{\mu}\gamma ^{5} + 
 F_{\mathrm P}q^{\mu}\gamma ^{5}\Big] {\cal O}_{\tau} \ ,
	     \label{eq.cc}
\end{eqnarray}
where ${\cal O}_{\tau} = \tau^{\pm}$ are the isospin operators for 
charged-current (CC)
reactions, while ${\cal O}_{\tau} = 1$ for neutral-current (NC) scattering. The 
induced pseudoscalar form factor $F_{\mathrm P}$ contributes only to CC 
scattering but its 
effects are almost negligible.
For NC reactions, the weak isovector form factor, $F_1^{\mathrm V}$ and 
$F_2^{\mathrm V}$, and the axial
form factor are expressed as in Eqs. (\ref{eq.fnc}) and (\ref{eq.gas}),
whereas for CC scattering they are
\inieq
 F_i^{\mathrm V} = F_i^{\mathrm p} - F_i^{\mathrm n} \ , \ \ \ 
 G_{\mathrm A} =   g_{\mathrm A} G \ 
, \label{eq.fcc} \fineq
where $g_{\mathrm A} \simeq 1.26$ and $G$ is defined in Eq.~(\ref{eq.ga}).
The single differential cross section for the quasi-elastic 
$\nu$($\bar\nu$)-nucleus scattering with respect to the outgoing nucleon 
kinetic energy $\mathrm {T_N}$ is obtained after performing an integration 
over the solid 
angle of the final nucleon and over the energy and angle of the final lepton.
We use in our calculations a relativistic optical potential with a real and 
an imaginary part which produces an absorption of flux. 
This is correct for an exclusive reaction, but would be
incorrect for an inclusive one. Here we consider situations where an emitted 
nucleon is
detected and treat the quasi-elastic neutrino scattering as a process where the
cross section is obtained from the sum of all the integrated exclusive 
one-nucleon knockout channels.
Some of the reaction channels which are
responsible for the imaginary part of the optical potential, like multi-step 
processes, fragmentation of the nucleus, absorption, etc. are not included in 
the experimental cross section as an emitted proton is always detected. 
The outgoing proton can be re-emitted after re-scattering 
in a detected channel, thus simulating the kinematics of a quasi-elastic 
reaction, only in few cases. The relevance of this contributions depends 
on kinematics and should not be too large in the situations considered here.
\begin{figure}[h]
\begin{center}
\includegraphics[width=3.5in]{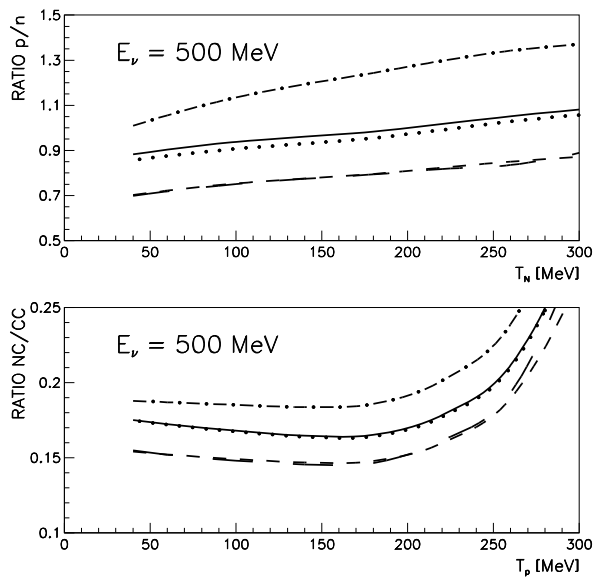}   
\vskip -0.3cm
\caption {Upper panel: ratio of proton-to-neutron NC cross sections of the 
$\nu$ scattering
on $^{12}$C . Lower panel: ratio of neutral-to-charged current cross sections 
of the $\nu$ scattering on $^{12}$C. Dashed lines are the results with no 
strangeness contribution, 
solid lines with $g^{\mathrm s}_{\mathrm A} = -0.10$, dot-dashed lines
with $g^{\mathrm s}_{\mathrm A} = -0.10$ and $\mu^{\mathrm s} = -0.50$,
dotted lines with $g^{\mathrm s}_{\mathrm A} = -0.10$ and 
$\rho^{\mathrm s} = +2$.
Long dashed lines are the RPWIA results without strangeness contribution.
}
\label{3f}
\end{center}
\end{figure}

Since an absolute cross section measurement is a very hard experimental task 
due to difficulties in the determination of the neutrino flux
in Ref.~\cite{gar92}  was suggested to measure the ratio of proton to 
neutron (p/n) yield as an alternative way to separate the effects of the 
strange-quark contribution. This ratio is 
very sensitive to the strange-quark contribution as the axial-vector
strangeness $g^{\mathrm s}_{\mathrm A}$ interferes with the isovector
contribution $g_{\mathrm A}$ with one sign in the 
numerator and with the opposite sign in the denominator 
(see Eq.~(\ref{eq.gas})) and it 
is expected to be less sensitive to distortion effects than the cross sections
themselves. 
In the upper panel of Fig.~\ref{3f} 
the p/n ratio  of the quasi-elastic $\nu$ scattering on $^{12}$C is displayed 
as a function of $\mathrm {T_N}$. The RPWIA results 
are shown in the figure and they are almost coincident with RDWIA ones. The 
p/n ratio for an incident neutrino is enhanced by a factor 
$\simeq 20$-$30$\% when 
$g^{\mathrm s}_{\mathrm A}$ is included and by $\simeq 50$\% when both  
$g^{\mathrm s}_{\mathrm A}$ and $\mu^{\mathrm s}$ are included. 
A minor effect is produced by $\rho^{\mathrm s}$, which gives 
only a slight reduction of the p/n ratio. 
Precise measurements of the p/n 
ratio appear however problematic due to the difficulties associated with 
neutron detection. This is the reason why
the most attractive quantity to extract experimental information about
the strangeness content seems the ratio of the neutral-to-charged (NC/CC) cross 
sections. 
In fact, although sensitive to the strange-quark effects only in the numerator, 
the NC/CC ratio is simply related to the number of events with an outgoing
proton and a missing mass with respect to the events with an outgoing proton in
coincidence with a muon. 
Our RDWIA results for the NC/CC ratio 
of the quasi-elastic $\nu$ scattering on $^{12}$C 
are presented in the lower panel of Fig.~\ref{3f} 
as a function of the kinetic energy of the outgoing proton.
The fact that the CC cross section goes to zero more rapidly
than the corresponding NC one (because of the muon mass) causes the 
enhancement of the ratio at large values of T$_{\mathrm p}$. The simultaneous 
inclusion of 
$g^{\mathrm s}_{\mathrm A}$ and $\mu^{\mathrm s}$ gives an enhancement that is 
about a factor of 2 larger than the one corresponding to the case with
only $g^{\mathrm s}_{\mathrm A}$ included. The effect of $\rho^{\mathrm s}$ is
very small.


\begin{thebibliography}{}


\bibitem{emc}
J. Ashman, {\it et al.}, [European Muon Collaboration], {\em Nucl. Phys. B}
{\bf 328}, 1 (1989).

\bibitem{spa2}
D.T. Spayde, {\it et al.},
{\em Phys. Lett. B} {\bf  583}, 79 (2004).

\bibitem{happex2}
K. Aniol, {\it et al.}, [HAPPEX Collaboration],
{\em Phys. Rev. Lett.} {\bf 82}, 1096 (1999);
{\em Phys. Rev. C} {\bf 69}, 065501 (2004); 
{\em Phys. Lett.  B} {\bf 635}, 275 (2006).


\bibitem{a42}
F.E. Maas, {\it et al.}, [A4 Collaboration],
{\em Phys.\ Rev.\ Lett.\ } {\bf 94}, 152001 (2005). 

\bibitem{g02}
D.S. Armstrong, {\it et al.}, [G0 Collaboration],
{\em Phys.\ Rev.\ Lett.}  {\bf 95}, 092001 (2005). 

\bibitem{happex24}
A. Acha, {\it et al.}, [HAPPEX Collaboration], 
nucl-ex/0609002.

\bibitem{bnl}
L.A. Ahrens, {\sl et al.},
{\em Phys. Rev. D}  {\bf 35}, 785 (1987).

\bibitem{gar} 
G.T. Garvey, {\it et al.},
 {\em Phys. Rev. C} {\bf 48}, 761 (1993).

\bibitem{fin}
S. Brice, {\it et al.}, hep-ex/0402007. 
See also \url{http://www-finesse.fnal.gov/index.html} and 
\url{http://www-boone.fnal.gov/}.

\bibitem{pate}
S.F. Pate,
{\em Phys. Rev. Lett.} {\bf 92}, 082002 (2004);
S.F. Pate, G. MaLachlan, D. McKee, and V. Papavassiliou, 
hep-ex/0512032.

\bibitem{pv}
A. Meucci, C. Giusti, and F.D. Pacati, 
{\em Nucl. Phys. A} {\bf 756}, 359 (2006).
 
\bibitem{bernard} 
V. Bernard, L. Elouadrhiri, and Ulf-G. Meissner,
 {\em J. Phys. G}  {\bf 28}, R1 (2002).

\bibitem{ee}
A. Meucci, F. Capuzzi, C. Giusti, and F.D. Pacati,
 {\em Phys. Rev. C} {\bf 67}, 054601 (2003).

\bibitem{cc}
A. Meucci, C. Giusti, and F.D. Pacati, 
{\em Nucl. Phys. A} {\bf 739}, 277 (2004).

\bibitem{sharma}
M.M. Sharma, M.A. Nagarajan, and P. Ring,
{\em Phys. Lett. B} {\bf 312}, 377 (1993).

\bibitem{chc} 
E.D. Cooper, {\it et al.},
{\em Phys. Rev. C} {\bf 47}, 297 (1993).

\bibitem{lala}
G.A. Lalazissis, J. K\"onig, and P. Ring, 
{\em Phys. Rev. C} {\bf 55}, 540 (1997).

\bibitem{beck1}
D.H. Beck and R.D. McKeown,
{\em Ann. Rev. Nucl. Part. Sci.} {\bf 51}, 189 (2001).

\bibitem{book}
S. Boffi, {\it et al.},
{\it Electromagnetic Response of Atomic Nuclei}, Oxford Studies in Nuclear
Physics, Vol. 20 (Clarendon, Oxford, 1996);
S. Boffi, {\it et al.}, {\em Phys. Rep.} {\bf 226}, 1 (1993).

\bibitem{nc}
A. Meucci, C. Giusti, and F.D. Pacati, 
{\em Nucl. Phys. A} {\bf 773}, 250 (2006).

\bibitem{meucci1}
A. Meucci, C. Giusti, and F.D. Pacati, 
 {\em Phys. Rev. C} {\bf 64}, 014604 (2001);
{\em Phys. Rev. C}  {\bf 64}, 064615 (2001).

\bibitem{gar92}
G.T. Garvey, {\it et al.},
{\em Phys. Lett. B} {\bf 289}, 249 (1992); 
G.T. Garvey, {\it et al.},
 {\em Phys. Rev. C}  {\bf 48}, 1919 (1993);
C.J. Horowitz, {\it et al.}, {\em Phys. Rev. C}  {\bf 48}, 3078 (1993); 
W.M. Alberico, {\it et al.},
{\em Nucl. Phys. A} {\bf 623}, 471 (1997); 
W.M. Alberico, {\it et al.}, {\em Phys. Lett. B} {\bf 438}, 9 (1998).




\end{thebibliography}
\end{document}